\documentclass[twocolumn,superscriptaddress,secnumarabic,amssymb,nobibnotes,nofootinbib,aps,pre,showpacs]{revtex4}
\usepackage{graphicx}
\begin{document}
\title{Wavelet analysis of event by event fluctuations}
\author{P. Manimaran$^\dag$\thanks{ph01ph11@uohyd.ernet.in} and
Prasanta K. Panigrahi$^\ddag$\thanks{prasanta@prl.ernet.in}}
\address{$^\dag$ School of Physics,
University of Hyderabad, Hyderabad, 500 046, India \\
$^\ddag$ Physical Research Laboratory, Ahmedabad, 380 009, India}
\date{\today}

\begin{abstract}

The temporal fluctuations of produced hadron density in heavy ion
collisions, modelled by 2D Ising model at temperatures $T_c$ and
below, are studied through a recently developed wavelet based
fluctuation analysis method. At $T_c$, long-range correlated
multifractal behavior, matching with the recently observed Hurst
exponent $H\simeq 1$, is found. Below $T_c$ uncorrelated
monofractal behavior is seen. The correlation behavior compares
well with the results obtained from continuous wavelet based
average wavelet co-efficient method, as well as with Fourier power
spectral analysis.

\end{abstract}


\maketitle

\section{Introduction}

The study of correlations and scaling behavior is an area of
active research \cite{mandel}. Various methods of analysis like
rescaled range analysis \cite{hur}, structure function
\cite{feder}, wavelet transform modulus maxima \cite{arn1},
detrended fluctuation analysis and its variants are used to study
the correlation behavior and fractal characteristics
\cite{gopi,ple,khu,chen,matia,krs,phand,xu}. Recently, we have
developed a new method based on discrete wavelet transform to
study the scaling properties of non-stationary time series
\cite{mani2}. This procedure economically extracts the
multifractal behavior in a time series. It has been applied to
characterize current and voltage fluctuations in tokamak plasma
and financial time series \cite{mani1}.

The goal of the present work is to apply this discrete wavelet
based approach for the study of event by event fluctuations of
hadron density produced during phase transition. The fact that,
our method is a local one, makes it ideal for the analysis of
fluctuations in a non-stationary data. The primary motivation for
this work is the recent study of the event by event fluctuations
of hadron multiplicities, carried out by Qin and Ta-chung,
\cite{liu1,liu2} using rescaled range analysis, which has shown
scaling behavior. The Hurst exponent $H$, has been found to be
$1$. The geometry of the hadronization scenario in the context of
heavy ion collisions, with two flavors, has led to the description
of the same through a model like two dimensional (2D) Ising model.
Hwa and co-workers have analyzed the scaling behavior of hadronic
fluctuations through various methods like moment analysis etc., in
the context of the above model \cite{hwa0,hwa1,hwa2,hwa3}.

A large number of studies have been done to analyze the
characteristic behavior of hadron formation during quark hadron
phase transition. Earlier studies regarding the fluctuations in
particle production, like correlation analysis of hadronic
fluctuations have been carried out using cluster expansion
technique of Ursell and Mayer \cite{urs}. Bialas and Peschanski
have studied multiparticle production during quark hadron phase
transition through factorial moment analysis \cite{bia}. Using
statistical methods, Ludlam and Slansky studied the mechanism of
particle production during quark hadron phase transition
\cite{lud}.

Following Hwa et. al. \cite{hwa2}, we have carried out the
Monte-Carlo simulation of 2D Ising model to mimic the
multi-particle production during quark hadron phase transition.
The time series of event by event fluctuations of average hadron
densities below and at phase transition have been measured for
characterization. The spatio-temporal fluctuations of produced
hadrons show correlation behavior which have been characterized
using wavelet transform techniques. We have made use of discrete
and continuous wavelet transforms to analyze the hadronic
fluctuations at various temperatures below and at critical
temperature. We also study the scaling behavior through power
spectral analysis.

\begin{figure}
\centering
\includegraphics[width=3in]{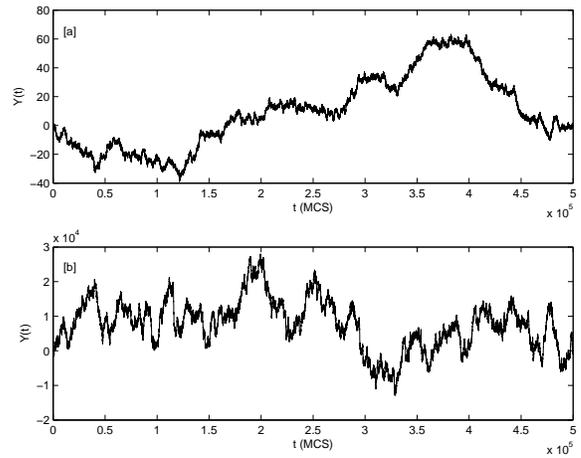}
\caption{The integrated time series of average hadron densities
after subtraction of the mean, [a] at $T = 1.0$, below $T_c$ and
[b] $T_c = 2.27$. The difference in behavior of the fluctuations
at different temperatures is clearly visible.}
\end{figure}

Wavelet transform is a mathematical tool which finds applications
in various fields, ranging from finance \cite{meg,mani1}, biology
\cite{pkp1,pkp2}, to chemistry \cite{chu}, and physics
\cite{arn1,ding,mani0} etc. Wavelet transform decomposes any given
function or data in a manner, which achieves optimal
time-frequency localization. Wavelets are classified into discrete
and continuous ones \cite{daub,mall}. We have made use of both the
methods to analyze the fluctuations in this paper. In case of
discrete wavelet analysis, we have used an approach which has been
developed recently \cite{mani2}. Average wavelet coefficient
method, a continuous wavelet approach as well as Fourier analysis
are also used to study the correlation behavior of the
fluctuations and corroborate our findings.

This paper is organized as follows. Section II deals with hadron
density as computed from the Ising model. In Section III, discrete
wavelet analysis of time series has been carried out for
extracting multifractal behavior. In section IV, continuous
wavelet analysis and Fourier power spectral analysis have been
implemented to support the previous discrete wavelet transform
approach. Section V concludes with discussions and future
directions of work.

\section{Hadron density and the Ising model}
In the context of heavy ion collisions, a two dimensional geometry
can be invoked, where 2D Ising model captures the dynamics of the
phase transition in a two flavor scenario. Earlier studies by Hwa
et.al., \cite {hwa2} describe in detail the relationship between
hadron density and the 2D Ising model. In this paper, we adopt the
same procedure and deal with the correlation analysis and fractal
characteristics through wavelet transforms. We have studied the
hadronic fluctuations using both discrete and continuous wavelet
transform as a time series analysis.

The two dimensional Ising model is defined by the Hamiltonian,
\begin{equation}
H = - J \sum_{\langle i,j \rangle} s_i s_j
\end{equation}
where $s$ takes the values $\pm 1$ and the sum is over the nearest
neighbors. In our Monte-Carlo simulation of 2D Ising model
\cite{landau}, we use Wolff single cluster algorithm \cite{wolff}
for spin flipping in a configuration of a lattice size $L = 256$.
We define the cell size to be $\epsilon^2$ with $\epsilon = 4$ and
the whole lattice is subdivided into $l=(L/\epsilon)^2$ cells. The
hadron density at a particular cell is defined by
\begin{equation}
 \rho_{i}= \lambda\ c_i^2
 \Theta\left(c_i\right),
\end{equation}
where $c_i$ is the net spin at cell $i$ defined to be positive
along the overall magnetization, i.e.,
\begin{equation}
 c_i= \left( sign \left( \sum_{j \epsilon L^2} s_j\right) \right )\sum_{j \epsilon A_i}
 s_j.
\end{equation}
$A_i$ is the cell block of size $\epsilon^2$ located at $i$,
$\Theta$ is the Heaviside function and $\lambda$ is a constant.
During the hadronization process $c_i$ fluctuates from cell to
cell, which reflects in the average hadron density in event to
event. The average density is calculated as;
\begin{equation}
<\rho>=\frac{1}{N} \sum_{e=1}^N \frac{1}{l} \sum_{i=1}^{l} \rho_i.
\end{equation}
Here, $i$ denotes the cells and $e$ is the configurations
captured, where $N$ is the number of events simulated. We have
calculated the equilibrium time series of the average hadron
density over $N=5$ x $10^5$ iterations.

\section{Wavelet based fluctuation analysis}

Considering the time series of average hadron density ($\langle
\rho \rangle_i$) from $i = 1,...,N$, we capture configurations at
various temperatures like below, and at critical temperature. In
most applications, the index $i$ will correspond to the time of
the measurements. The profile ($Y_i$) of the time series is
obtained by subtracting the mean and taking cumulative sum of the
average hadron density. The profiles have been analyzed by the
wavelet based fluctuation analysis method to study the scaling
behavior \cite{mani2}. We calculate the scaling exponents $h(q)$
for various moments $q$, in this case $q$ varies from $-10$ to
$10$. For $q=2$, $h(q)$ is the Hurst exponent, which is one of the
measures of correlation behavior in the time series analysis: $0
\leq H \leq 1$. For persistent time series $H > 0.5$, $H=0.5$ for
uncorrelated time series and $H < 0.5$ for anti-persistent
behavior. Recently, the anti-persistent regime $0 < H < 0.5$ has
been mapped to $\Delta_3$ statistics, widely employed in random
matrix theory \cite{santh}. The main aim here is to study the
fractal characteristics of the time series and its multifractal
nature. For the time series possessing multi-fractal behavior,
$h(q)$ decreases with increasing values of $q$. For mono-fractal
time series $h(q)$ is constant for all values of $q$. Fig. 1
depicts the profile of the time series of event by event
fluctuations of average hadron density, [a] below $T_c$, and [b]
at $T_c$.

We use a discrete wavelet based fluctuation analysis method (which
is analogous to multifractal detrended fluctuation analysis
method) to study the existence of scaling behavior of the time
series. Our earlier paper gives the detail procedure of this
approach \cite{mani2}. The fluctuations are extracted using a
wavelet (Db-8) from Daubechies family. The fluctuation function
$F_q (s)$ is obtained for various scales $s$. The power law
manifests itself as a straight line in the log-log plot of
$F_q(s)$ versus $s$, for each value of $q$.
\begin{equation}
F_q(s) \sim s^{h(q)}
\end{equation}
From the analysis of hadron density time series, below $T_c$,
uncorrelated monofractal behavior and long-range correlated
multifractal behavior at $T_c$ are found. These results are shown
in Fig. $2$. It was found that the Hurst exponent $H=h(q=2) \simeq
1$ matches with that of Qin et. al. \cite{liu}. Very
interestingly, multifractal behavior is clearly seen for event by
event fluctuations at $T_c$.

\begin{figure}
\centering
\includegraphics[width=3in]{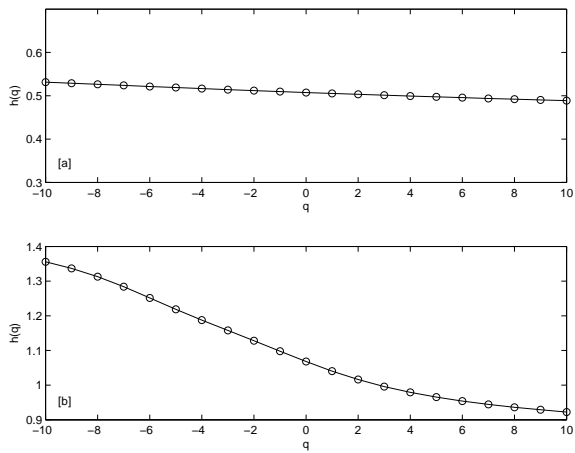}
\caption{For time series [a] Below $T_c$, $h(q)$ values shows
linear behavior for different values $q$ indicating the
monofractal behavior and [b] the non-linear behavior of $h(q)$
values for different values of $q$ at $T_c$, shows clearly the
long-range correlation and multifractal nature.}
\end{figure}

\section{Average wavelet coefficient method}

We now make use of average wavelet coefficient method to
corroborate the above findings. This is a continuous wavelet based
approach, which has been used to study the temporal correlations
of the fluctuations in time series. We obtain the Hurst measure by
transforming the time series, $f(t)$ into wavelet domain. The
continuous wavelet transform of a function $f(x)$ is given by

\begin{equation}
W[f](a,b) = \frac {1}{\sqrt a} \int_{- \infty}^{\infty}
\psi_{a,b}^{*}(x) f(x) dx,
\end{equation}
where the mother wavelet $\psi_{a,b}$ is defined as,
\begin{equation}
\psi_{a,b} (x) = \psi \left( \frac{x-b}{a} \right).
\end{equation}

Here $a$ is the scaling parameter $(a>0)$ and $b$ is the
translation parameter $(- \infty < b < \infty)$ and $\psi^{*}(x)$
is the complex conjugate of $\psi(x)$.

It has been shown that \cite{awc},
\begin{equation}
W[f](a) = \langle | W[f](a,b) | \rangle \simeq a ^{\frac{1}{2}+H}.
\end{equation}

\begin{figure}
\centering
\includegraphics[width=3in]{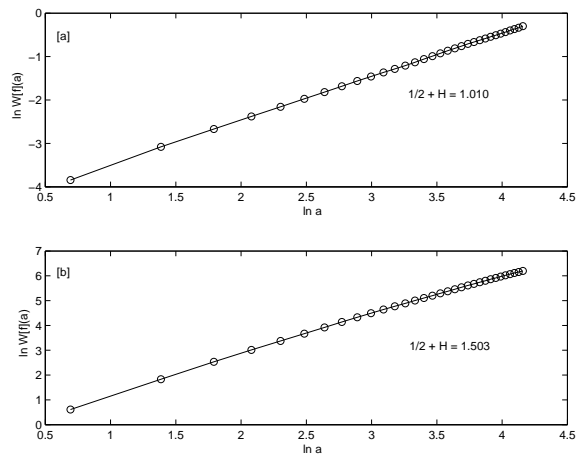}
\caption{Log-Log plot of average wavelet coefficients $W[f](a)$
versus scale $a$, indicates [a] below $T_c$, uncorrelated behavior
($H \simeq 0.5$) and [b] at $T_c$, long-range correlation ($H
\simeq 1$).}
\end{figure}

At a given scale, the wavelet energy of all locations have been
averaged yielding $W[f](a)$. The same averaging is followed for
all scales. Thus for a self-affine time series, the scaling
exponent ($\frac{1}{2}+H$) is measured from the slope of the
log-log plot of $W[f](a)$ versus scale $a$. The results are shown
in Fig. 3. They clearly indicate that the fluctuations possess
self-affine nature and the Hurst exponent is the measure of
correlation. From the obtained $H$, we found that the hadronic
fluctuations are uncorrelated below $T_c$ ($H =0.5$) and at $T_c$
it possesses long-range correlation ($H\simeq1$). This matches
with our previous discrete wavelet analysis.

We have also analyzed the scaling behavior through Fourier power
spectral analysis,
\begin{equation}
P(s) = \left | \int Y(t)~~exp( -2 \pi i s t)~~dt \right |^2.
\end{equation}
Here $Y(t)$ is the accumulated fluctuations after subtracting the
mean $\langle Y \rangle$. It is well known that,  $P(s) \sim s^{-
\alpha}$. The obtained scaling exponent $\alpha$ is compared with
Hurst exponent by the relation $\alpha = 2 H + 1$. For the time
series far below from $T_c$, the scaling exponent $\alpha = 2$
which reveals uncorrelated Brownian motion behavior and at $T_c$,
the scaling exponent $\alpha = 3$, which reveals long range
correlated fractional Brownian motion behavior as shown in Fig. 4.

\begin{figure}
\centering
\includegraphics[width=3in]{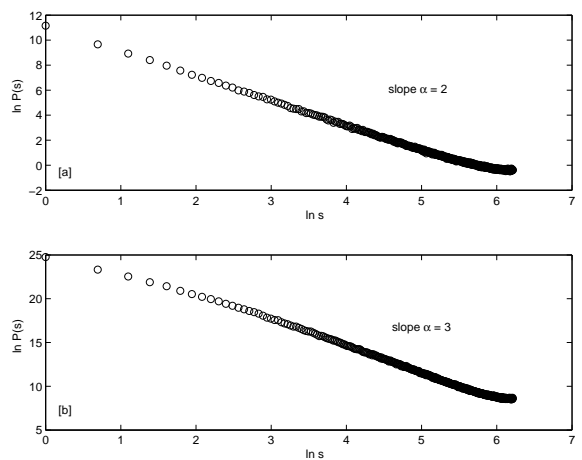}
\caption{Log-Log plot of Fourier power spectrum $P(s)$ versus
frequency $s$, shows [a] below $T_c$, Brownian motion behavior
$\alpha = 2$ and [b] at $T_c$, fractional Brownian motion $\alpha
= 3$.}
\end{figure}

\section{Conclusion}
We have found that the event by event fluctuations of hadron
density, as modelled by 2D Ising spins, possess long-range
correlated multifractal and fractional Brownian motion behavior at
$T_c$. This is due to the correlation arising from phase
transition. Uncorrelated monofractal and Brownian motion behavior
below $T_c$ is observed as per physical expectations. The analysis
has been carried out through discrete wavelet wavelet based
fluctuation analysis method, which is well suited for
non-stationary data. The continuous wavelet based average wavelet
coefficient method and Fourier analysis fully supported the above
conclusion.

We thank Prof. J. C. Parikh and M. S. Santhanam for useful
discussions.

\end{document}